\documentclass[12pt]{iopart}

\usepackage{graphicx}
\usepackage{amssymb}

\newcommand{\ii}{{\rm i}}
\newcommand{\dd}{{\rm d}}

\newcommand{\bC}{{\mathbb{C}}}

\newcommand{\bR}{{\mathbb{R}}}

\newcommand{\bT}{{\mathbb{T}}}

\newcommand{\ave}[1]{\left\langle #1 \right\rangle}
\newcommand{\mod} {\mathop{\rm mod}}

\newcommand{\beq}{\begin{equation}}
\newcommand{\eeq}{\end{equation}}
\newcommand{\beqa}{\begin{eqnarray}}
\newcommand{\eeqa}{\end{eqnarray}}
\newcommand{\beqas}{\begin{eqnarray*}}
\newcommand{\eeqas}{\end{eqnarray*}}

\begin{document}

\title{A hybrid method for calculation of Ruelle-Pollicott resonances}

\author{Martin Horvat}

\address{Department of Physics, Faculty of Mathematics and Physics, University of Ljubljana, Jadranska 19,  SI-1000 Ljubljana, Slovenia}

\address{CAMTP - Center for Applied Mathematics and Theoretical Physics, University of Maribor, Krekova 2, SI-2000 Maribor, Slovenia}

\eads{\mailto{martin.horvat@fmf.uni-lj.si}}

\author{Gregor Veble}

\address{University of Nova Gorica, Vipavska 13, P.P. 301 Ro\v zna Dolina, SI-5000 Nova Gorica, Slovenia}

\address{Pipistrel d.o.o. Ajdov\v s\v cina, Gori\v ska c. 50c, SI-5270 Ajdov\v s\v cina, Slovenia}

\address{CAMTP - Center for Applied Mathematics and Theoretical Physics, University of Maribor, Krekova 2, SI-2000 Maribor, Slovenia}

\eads{\mailto{gregor.veble@p-ng.si}}

\date{\today}

\begin{abstract}
We present a numerical method for calculation of Ruelle-Pollicott resonances of dynamical systems. It constructs an effective coarse-grained propagator by considering the correlations of multiple observables over multiple timesteps. The method is compared to the usual approaches on the example of the perturbed cat map and is shown to be numerically efficient and robust.
\end{abstract}

\pacs{05.45.-a,
      95.10.Fh, 
      76.20.+q, 
      05.45.Pq,
      02.60.-x
      }

\submitto{\JPA}


\section{Introduction}

The calculation of Ruelle-Pollicott resonances is a nontrivial problem in dynamical systems theory from both a theoretical as well as a numerical viewpoint. They describe the asymptotic decay of phase space densities towards the invariant phase space measure. The Perron-Frobenius (PF) operator gives the propagation of phase space densities, and the Ruelle-Pollicott resonances are the eigenvalues of the coarse-grained PF operator. The coarse-graining is a crucial element in their description. In a chaotic system, the stretching and folding of phase space structures causes the phase space densities to attain finer and finer structures under the evolution of a PF operator. It is only after the coarse graining that these structures are smeared out and that the decay towards the invariant measure can be observed.

Here we are concerned with numerical approaches towards the calculation of Ruelle-Pollicott resonances. There exist two typical approaches towards their calculation. The most common one is to express the PF operator in a finite basis that also provides a natural way of coarse graining, see e.g. \cite{blum, sano}. This reduces the problem to the one of finding the eigenvalues of a matrix. The problem with such an approach is that it requires large matrices in order for the resolution to be sufficient for good convergence. With increasing matrix size, however, the coarse graining is also reduced, causing the creation of a large number of spurious eigenvalues that, in the case of area preserving invertible maps, tend to fill out the whole unit circle. This makes even the identification of resonances a difficult task \cite{haake}. This can be considered as a many-observable, single-timestep approach. The numerical difficulties stem from the fact that, in order to obtain a good long-time description of the system, a large basis needs to be considered to faithfully capture the actual dynamics from a single timestep.

A different approach considers taking only a single observable (phase space function) and, instead of focusing on a single iteration of the PF operator, the auto-correlation function of such an observable is computed over many steps of the iteration. By using linear predictors or Pad\'e approximates the signal is fitted to a sum of exponentials, which in turn give the positions of the resonances, see e.g. \cite{isola}. A similar technique is used in \cite{stein, main} for determining the spectral lines of quantum systems. A whole class of numerical schemes based on this approach is presented in \cite{florido}. Compared to matrix diagonalization, this approach is computationally cheaper. The disadvantage of such an approach is a certain amount of arbitrariness when choosing the number of exponentials to be fitted to the signal. As will be shown, this can introduce significant uncertainties into the obtained results. Also, while such an approach faithfully captures the long time behaviour of the system, the results also depend on the behaviour of a single chosen observable. This might cause certain resonances to be missed or poorly resolved.

We present a hybrid method that combines the favourable aspects of the described approaches. It uses a moderate number of observables as well as a moderate number of timesteps when estimating the eigenvalues of the PF operator. It will be shown that such an approach requires fewer observables than the direct diagonalization technique, and it provides its own criterion on the optimal number of time steps. By using multiple simultaneous observables, it also avoids the problem of missing resonances.

In the following, we define the Perron-Frobenius operator and the Ruelle-Pollicott resonances. We then discuss the issues with the linear predictor approach and give an analytical estimate on the spurious "ring" of resonances that can cause numerical difficulties. The hybrid method is presented next, followed by the comparison of results for various methods, and conclusions.

\section{Preliminaries}

We consider an ergodic and mixing \cite{ott} time-discrete dynamical system $f^t:{\cal S} \to {\cal S}$  in the phase-space ${\cal S}$ with the invariant measure $\mu:{\cal S} \to \bR_+$  normalized so that $\mu({\cal S})=1$. A single time step of the system is defined by the iteration 
\beq
  x_{t+1} = f(x_t)\>,\qquad x_t \in {\cal S}\>.
\eeq
Instead tracking individual trajectories the dynamics can be expressed by the evolution of the their probability distribution over the phase space $\rho^t:{\cal S} \to \bR_+$ defined using the Perron-Frobenius (PF) operator $L$ as
\beq
  \rho_{t+1}(x) = (L \rho_t)(x)= \int \dd y\, \delta(x-f(y))\rho_t(x)\>.
\eeq
This can can be explicitly written in the form of a sum over all $y$ that are mapped into $x$ by the map $f(\cdot)$ reading
\beq
  (L \rho_t)(x) = \sum_{y \in f^{-1}(x)} \frac{\rho_t(y)}{|\det(f^\prime(y))|}\>,
\eeq
where $f^\prime(x)$ is the Jacobian of the map. For area preserving invertible maps the invariant measure is Lebesgue $\dd\mu(x)=\dd x/\int_{\cal S} \dd y$ and the action of the PF operator reduces to
\beq
  \rho_{t+1}(x)  = (L \rho_t)(x) = \rho_t(f^{-1}(x))\>.
\eeq
Let us denote by $\mu(g)$ the phase-space average $\int\dd \mu(x) g(x)$ of a function $g(x)$ w.r.t. to the invariant measure $\mu(\cdot)$. Then the correlation of two real smooth observables $u(x)$ and $v(x)$, with their phase-space averages $\mu(u)$ and $\mu(v)$ equal to 0, is given by
\beq
  C_{u,v} (t) = \mu (u \cdot L^t \circ v)\>.
\eeq
In mixing systems and under certain conditions \cite{ruelle,pollicott}, it has a well defined asymptotic time dependence of the form
\beq
  C_{u,v} (t) \sim \Re\{A_{u,v}\,\nu^t\} \>,
\eeq
where $A_{u,v}$ is a constant depending on used observables and $\nu$ is called the leading Ruelle-Pollicott (RP) resonance. The RP resonances are the eigenvalues of the coarse grained PF operator \cite{gaspard}. Here we are interested mostly in dynamical systems confined to a compact phase space in which the RP resonances are believed to depend on the fine structure of the mixing process as opposed to the diffusion in phase space. Namely, for a large area preserving maps on the cylinder (an infinite phase-space) it was shown in \cite{venegeroles07,venegeroles08} that there exists a strict connection between the resonances and the diffusion process. 

\section{Discussion of the linear predictor approach to Ruelle-Pollicott resonances}
\label{sec:lp}

A common type of approach towards calculating the resonances of arbitrary signals is to use linear predictors, which match the signal (or rather its auto-correlation function) to a sum of exponentials \cite{isola,stein, main,florido}.  If we consider a dynamical system with the PF operator $L$, then in the linear prediction (LP) approach we assume that an auto-correlation function $C(t)$ of some observable $u(x)$ with a zero phase-space average $\mu(u)=0$ can, for $t\ge 0$, be represented by a finite sum of exponentials as
\beq
  C_u(t)=\mu(u\cdot L^t u) \approx \sum_{i=1}^r C_i \nu_i^t \>,
  \label{eq:lin_expansion}
\eeq
where $\{\nu_i\}_{i=1}^r$ are the LP resonances and $C_i\in \bC$ are constant weights associated with the individual resonances.  In an ideal case, the LP resonances represent or approximate the RP resonances of the dynamical system. Equivalently, the auto-correlation signal $C(t)$ can approximated by a finite linear predictor (LP) \cite{makhoul, vaidyanathan}, which is given in the form of a finite difference recursion 
\beq
  C(t) = \sum_{i=1}^r d_i C(t-i) + \xi (t) \>,\qquad t = r,r+1,\ldots\>,
  \label{eq:lin_pred}
\eeq
with a discrepancy $\xi_t$ between the value produced by the finite LP model and the true value of the auto-correlation function $C(t)$. There are several known methods to obtain LP coefficients $d_i$ \cite{makhoul}, from which the least-squares is the most straightforward and therefore used here. 

For a given auto-correlation signal $\{ C(t)\}_{t=0}^n$ of length $n$, the vector of LP coefficients  $d=(d_i)_{i=1}^r$ is, in the least-squares approach, chosen such that the total discrepancy $\sum_{t=r}^n \xi(t)^2$ is minimal. This condition yields a linear system of equations for the vector $d$ written as
\beq
  A\, d = b\>,
  \label{eq:lp_LS}
\eeq
with the matrix $A= [A_{i,j}]_{i,j=1}^r$ and the vector $b=[b_i]_{i=1}^r$ given by
\beq
\fl\qquad
   A_{i,j} = \sum_{t=0}^{n-r} C(t+r-i)C(t+r-j)\>,\quad
   b_i = \sum_{t=0}^{n-r} C(t+r-i)C(t+r)\>.
  \label{eq:lp_LS_elem}
\eeq
This system of equation is usually well conditioned and the solution -- the vector $d$ -- determines the form of the LP. The LP resonances $\{\nu_i\}_{i=1}^r$ are given as the roots of the characteristic polynomial 
\beq
  P_{\rm LP}(x) = x^r - \sum_{i=1}^r d_i x^{r-i}
\eeq
corresponding to the LP. \par
\begin{figure}[!htb]
\centering
\includegraphics[width=12cm]{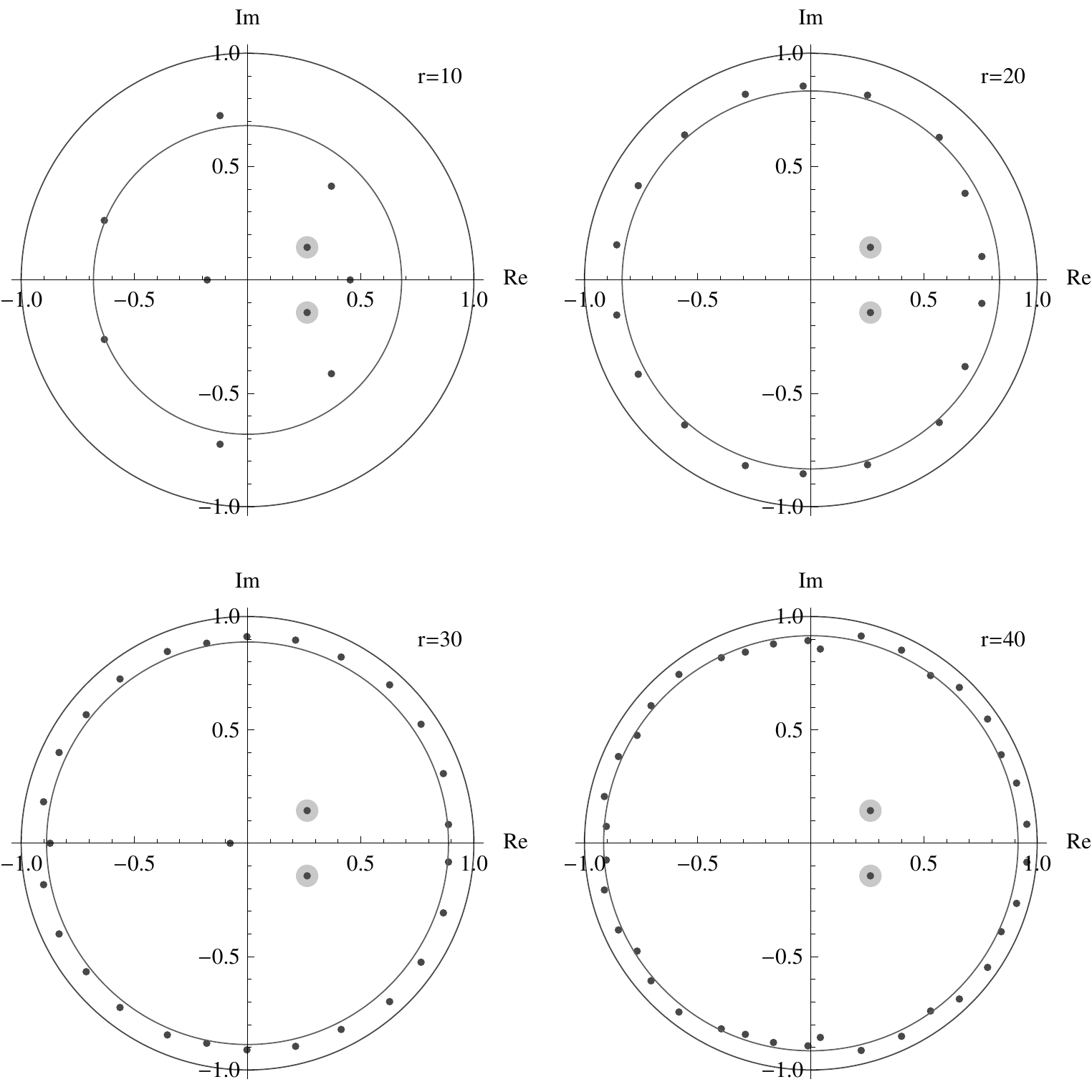}
\caption{The spectrum of LP resonances corresponding to the the auto-correlation $C(t)= \Re\{\nu^t\}+ \epsilon(t)$ for a different number of considered resonances $r$ with $\nu=0.3 e^{\ii\, 0.5}$, $t\in[0,n=10^3]$ and $\epsilon(t)$ being random and uniformly distributed in $[-10^{-12}, 10^{-12}]$. The radius of the circle going through the ring of resonances has been estimated by $R(r-2,n)$ (\ref{eq:ring_ana}). The gray-shaded discs mark the positions of $\nu$ and $\nu^*$.}
\label{pic:lin_sp}
\end{figure}

The number of exponential terms $r$ chosen is a critical parameter in such approaches. By choosing too few exponentials, the positions of the resonances are not properly resolved. Here we will show that, on the other hand, by taking too many exponentials, noise causes the fitting procedure to try and follow a random signal, which pushes the calculated resonances out towards the unit circle. We demonstrate this by considering one realization of a typical auto-correlation function of a real observable in system with only two clearly distinguishable complex resonances $\nu$ and $\nu^*$ written as
\beq
  C(t)=\Re\{\nu^t\}+\epsilon(t) \>,\qquad t=0,1,2,\ldots\>.
  \label{eq:oneres_corr}
\eeq
The function $\epsilon(t)$ represents a noise resulting from a finite precision of the auto-correlation measurement or calculations. In the case $r>2$ and $\epsilon(t)=0$ the auto-correlation possesses only two non-zero LP resonances $\nu$ and $\nu^*$.  The LP resonances corresponding to one realization of the auto-correlation (\ref{eq:oneres_corr}) with noise is depicted in figure \ref{pic:lin_sp}. 
We may see that the true resonances -- the complex conjugated pair $\nu$ and $\nu^*$ are present in the set of LP resonances. Beside the pair, there is a ring of resonances which is moving towards the unit circle as the number of resonances $r$ is increased. The ring of resonances is present whenever the number of considered resonances $r$ exceeds the number of detectable resonances.  In the following we will present a model giving a qualitatively correct radius of the ring of LP resonances.\par
Considering a LP model using $r$ resonances, we see that the ring radius composed of $q$ resonances is virtually independent from the amplitude of $r-q$ detectable resonances. We expect this ring to occur due to the noise present in the calculation of the correlation functions. This radius can therefore be expected to be estimated from the LP resonances corresponding to a numerical noise auto-correlation model $C_{\rm model}(t) = \delta_{t,0} + \epsilon(t)$ for $t\in[0,n]$ using $q$ resonances, where we assume that $\epsilon(t)$ is random white noise with variance $\upsilon=\ave{\epsilon(t)^2}$ much smaller than 1. 
By inserting $C_{\rm model}(t)$ into equations (\ref{eq:lp_LS_elem}) and taking into account the central limit theorem \cite{feller}, we find that the matrix elements $A_{i,j}$ and the vector elements $b_i$ are, in the limit $n\gg 1$, approximately Gaussian variables with statistical moments
\beq
\fl\qquad
  \ave{A_{i,j}} = \delta_{i,j} [\delta_{i,q} + \upsilon(n-q+1)]\>,\quad
  \ave{b_i} = 0\>,\quad \ave{b_i^2} = \upsilon [\delta_{i,q} + \upsilon (n-q+1)]\>,
\eeq
where $i,j\in[1,q]$ and the brackets $\ave{\cdot}$ represent the averaging over different noise realizations. The average of the matrix $A$ is diagonal and by neglecting the off-diagonal elements one can approximate the LP coefficients by
\beq
  d_i \approx \frac{b_i}{A_{i,i}} \>,
\eeq
which means that $d_i$ are approximately Gaussian variables with a zero average $\ave{d_i}=0$ and variances equal to
\beq
  \ave{d_i^2} = \frac{ \upsilon}
                {\delta_{i,q}+\upsilon(n-q+1)}\>.
\eeq
In the limit of small noise $\epsilon(t)\to 0$, the coefficient $d_q=0$, and the remaining non-zero LP coefficients $(d_1,\ldots,d_{q-1})$ define a characteristic random polynomial
\beq
  P_{\rm LP, model} \approx x\left ( x^{q-1} - \sum_{i=1}^{q-1} d_i x^{q-i} \right)
\eeq
corresponding to different realizations of the auto-correlation $C_{\rm model}(t)$. \par
\begin{figure}[!htb]
\centering
\includegraphics[width=8cm]{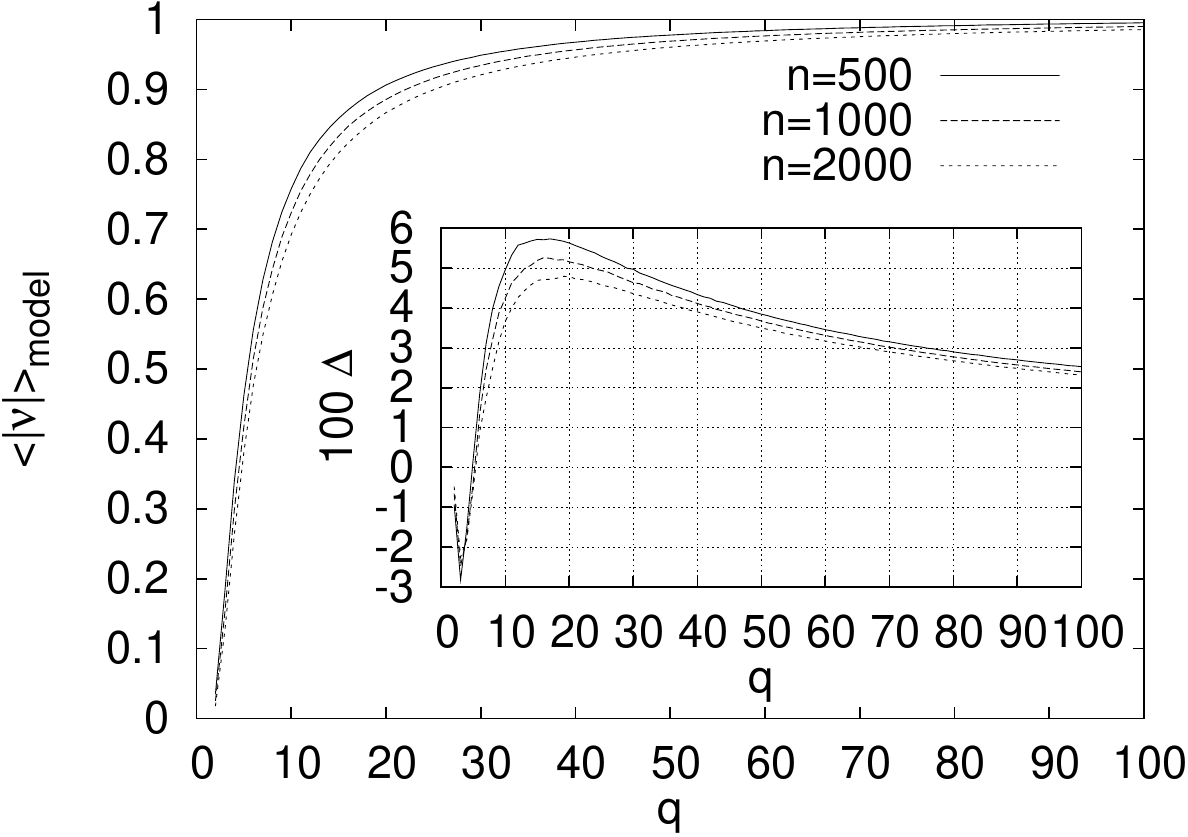}
\caption{The average amplitude of the largest LP resonance of the model auto-correlation $C(t)=\delta_{0,t}+ \epsilon(t)$ for different number of considered resonances $q$ and lengths $n$ of the auto-correlation signal. The inset shows the difference $\Delta$ between the average amplitude of the model and their analytic estimate (\ref{eq:ring_ana}). The $\epsilon(t)$ is random and uniformly distributed $[-10^{-12},10^{-12}]$. }
\label{pic:lin_ana}
\end{figure}

We find empirically that the roots of the polynomials $x^s-\xi \sum_{i=0}^{s-1} a_i x^i$ of order $s$, with coefficients $a_i$ being normalized Gaussian variables $N(0,1)$ and $|\xi|\ll 1$, form rings in the complex plane with the radius approximately $\xi^{\frac{1}{s}}$. This radius can also be obtained if we consider the zeros of a polynomial containing only the highest and constant order terms of the original polynomial, $x^s-\xi=0$, with the constant term replaced by its root mean squared value. According to this rather ad-hoc expression, the radius of resonances corresponding to $C_{\rm model}(t)$ is approximately given as
\beq
  R(q,n) \approx (n-q+1)^{-\frac{1}{2(q-1)}}\>.
  \label{eq:ring_ana}
\eeq
The comparison of this estimate with the average amplitude of the largest LP resonance is shown in figure \ref{pic:lin_ana}. We see that the analytic estimate is not very accurate, but it gives a qualitatively correct functional dependence. Similar estimates for the average radius of the ring of resonances in the complex plane can also be found for other approaches to obtain the LP coefficients, such as the auto-correlation method (sometimes called the Yule-Walker equation), the Burg's method etc.

While having the benefit of detecting the asymptotic behaviour directly, the linear predictor method is therefore disadvantaged by spurious solutions that make it hard to distinguish the real resonances from noise in practice. Furthermore, by focusing on a single observable, the results may significantly depend on the choice of the observable, which is nicely demonstrated in \cite{courbage}. On the other hand the LP methods are fast. The time complexity of preparing the data for the LP methods is $O(n)$, while calculating the LP resonances in a robust way takes $O(r^3)$ times steps.

\section{The hybrid method of detecting Ruelle-Pollicott resonances}

Here we propose a hybrid method that relies on the diagonalization of relatively small matrices, yet takes into account the information of multiple iterations of the PF operator. A somewhat analogous approach was taken in \cite{khodas} in the analytic exploration of the resonance structure of the kicked rotor. 

We start with the same approach as in the case of the direct diagonalization, but instead of a single matrix we compute a time-sequence of matrices. Let us choose a basis ${\cal B}_r = \{\varphi_i\}_{i=1}^r$ for the densities, which is orthonormal w.r.t. to the invariant measure $\mu(\varphi_i \cdot \varphi_j) = \delta_{i,j}$. The elements of the matrix corresponding to the $n$-th iteration of the map is given by
\beq
  T_{ij}^{(t)} = \mu \left(\varphi_i^{*}\cdot L^t \circ \varphi_j\right) \>. 
  \label{eq:correlationmatrices}
\eeq
One may also interpret this as a matrix of various cross correlations. In the usual diagonalization approach, one takes the matrix $T^{(1)}$ and computes its eigenvalues to obtain the resonances. In the linear prediction and related approaches, the diagonal matrix elements $T_{ii}^{(t)}$ are taken and fitted with a sum of exponentials as a function of $t$. 

Our approach is based on the principle that the iterations of the exact PF operator are related by 
\beq
  L^{t+1}=L\circ L^t\>.
  \label{eq:PFdecomp}
\eeq
The subsequent matrices $T^{(t)}$ should therefore be related by a single $t$-independent matrix $Y$ in an approximate way,
\beq
  T^{(t+1)}\approx Y T^{(t)}\>. 
  \label{eq:propagatorapproximation}
\eeq
The goal is to find a procedure to calculate the matrix $Y$ such that the above relation is best satisfied for all $t$. The eigenvalues of such a matrix are then expected to give the RP resonances. 
A way to calculate an optimal $Y$ is to minimize the error norm
\beq
  E(Y,Y^{\dag})
  =\sum_{t=0}^{n-1} w_t \Tr 
 \left [\left(T^{(t)\dag}Y^{\dag} - T^{(t+1)\dag}\right)
  \left(Y T^{(t)} -T^{(t+1)}\right) \right]\>. 
  \label{eq:error}
\eeq
The weights are introduced in order to boost the information from the later time steps, which provide the most information about the asymptotic decay, but would otherwise be drowned by the much larger correlations found in the first few timesteps. The choice is
\beq
  w_t = \left[\Tr\left(T^{(t)\dag} T^{(t)}\right)-\gamma \right]^{-1}\>,
  \label{eq:weight}
\eeq
where $\gamma=0$ if the basis set does not includes the uniform density and $\gamma=1$ if the basis set can fully reproduce the uniform density, as this is a trivial contribution that can be subtracted. With such a choice, information from all the time steps is treated with roughly the same significance. 

We may treat the matrix elements of the matrix $Y$ and its adjoint as independent complex numbers. We then try and minimize $E$ with respect to all the matrix elements by setting its gradient to $0$. One may note that
\beq
  \frac{\partial}{\partial A_{ij}}\Tr (AB)
  =\frac{\partial}{\partial A_{ij}}\sum_k \sum_l A_{kl} B_{lk}=B_{ji}\>.
\eeq
The requirement that the gradient of $\Tr AB$ with respect to all elements of $A$ equals $0$ then reduces to the matrix equation $B=0$. As the trace operation is invariant with respect to cyclic permutations of matrix products, we rearrange the order of matrix multiplications in equation (\ref{eq:error}) to obtain
\beq
 \fl\quad
  E(Y,Y^{\dag})
  =\sum_{t=0}^{n-1} 
  w_t \Tr 
  \left[
    Y^{\dag}
    \left(Y T^{(t)} T^{(t)\dag} - T^{(t+1)}T^{(t)\dag}\right) + 
    T^{(t+1)\dag} \left(Y T^{(t)} -T^{(t+1)}\right)
\right]
\eeq
and equaling the gradient of $g$ with respect to the  elements of $Y^\dag$ to 0 then leads to the equation
\beq
  \sum_{t=0}^{n-1} 
  w_t \left(Y T^{(t)} T^{(t)\dag} - T^{(t+1)}T^{(t)\dag}\right)=0\>.
\eeq
Solving this for $Y$ yields
\beq
  Y = \left(\sum_{t=0}^{n-1} w_t  T^{(t+1)}T^{(t)\dag}\right)
      \left(\sum_{t=0}^{n-1} w_t T^{(t)} T^{(t)\dag}\right)^{-1}\>.
\eeq
The RP resonance can be identified as the (sub-)dominant eigenvalue of the matrix $Y$. \par
It should be noted here that the presented linear predictor class methods can be extended so that not only a single auto-correlation function signal is fitted by exponentials, but signals of cross correlations of a set of observables given by equation (\ref{eq:correlationmatrices}) are considered. In such an approach, if the predictor is only a single step one, a super-matrix $\hat Q$ is sought such that the relationship
\beq
  T_{ij}^{(t+1)}=\sum_{k,l} \hat Q_{\left\{ij\right\}\left\{kl\right\}} T_{kl}^{(t)}
  \label{eq:supermatrix}
\eeq
is best satisfied for all iterations. If matrices $T^{(t)}$ are considered as vectors, the eigenvalues of $\hat Q$ are then expected to give the resonances of the system. With $r$ giving the basis size of the system, we obtain $r^2$ eigenvalues, whereas the direct diagonalization and hybrid approaches towards the calculation of the RP resonances return $r$ eigenvalues. This suggests a certain amount of underdetermination of the problem. It is likely for this reason that, when tested, the approach as given in equation (\ref{eq:supermatrix}) proved very unstable with a lot of spurious eigenvalues.

Equation (\ref{eq:propagatorapproximation}) can be considered as a particular reduction of equation (\ref{eq:supermatrix}) by only taking a very specific connection between correlations from subsequent timesteps. This reduction reflects the actual underlying process as given by equation (\ref{eq:PFdecomp}) and eliminates the underdetermination of the problem.

The calculation of all the matrices $\{T^{(t)} \}_{t=0}^n$ to some precision is a lengthy process having the time complexity of at least $O(n r^2)$, but typically even more ($O(n r^3)$ for the two-dimensional example studied later) as the integration grid has to be refined for faster oscillating observables, 
whereas obtaining the effective propagator $Y$ from matrices $T^{(t)}$ and calculating its eigenvalue spectrum takes $O(r^3)$ steps. In all usual cases the time complexity of direct diagonalization $(n=1)$ and of the hybrid method $n>1$ is several orders larger than for the LP methods when using the same number of resonances and time steps considered.

\section{Results of the hybrid method}

The presented method for searching of the RP resonances was testing using an area preserving map -- the perturbed Arnold's cat map defined on a two-dimensional torus $\bT^2 =[0,1)^2$ \cite{basilio} with one iteration step given by
\beqa
  y^{t+1} &=& y^t + x^t - \frac{K}{2\pi} \sin(2\pi x^t)\quad (\mod~~ 1) \>, \\
  x^{t+1} &=& x^t + y^{t+1}\quad (\mod~~ 1) \>.
\eeqa
This map can be shown to be fully ergodic for $K\in[0,1]$. At $K=1$ it develops a marginally stable periodic orbit at the origin. We therefore expect some RP resonances of the system to approach unity in terms of their absolute value when approaching $K\to 1$.
We choose to work with the real Fourier basis defined as
\beq
  \varphi_n(x)
  =\left\{ 
    \begin{array}{lll}
    1 &:& n = 0\cr
    \sqrt{2}\cos(2\pi n x) &:& n = {\rm even }>0\cr
    \sqrt{2}\sin(2\pi n x) &:& n = {\rm odd }
    \end{array}
  \right.
\eeq
because they are smooth, point convergent and non-local. We use $p$ Fourier modes along each axis and so the functions on the torus are spanned by the functional basis
\beq
 {\cal B}_{p^2} = \{ 
  \varphi_{i,j}(x,y) = \varphi_i(x) \varphi_j(y) 
  \}_{i,j=0}^{p-1} \>.
\eeq
Notice that the dimension of the vector space that we work with is $r=p^2$.\par
\begin{figure}[!htb]
\centering
{\footnotesize (a)}\hskip-5mm\includegraphics[width=7cm]{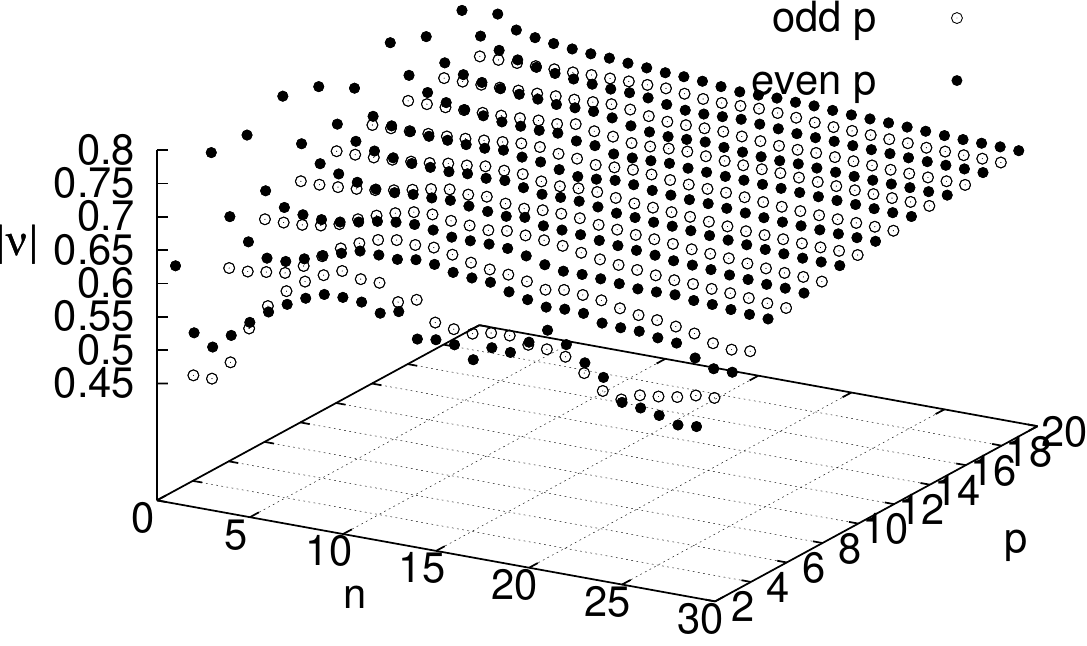}%
\hskip10mm{\footnotesize (b)}\hskip-5mm\includegraphics[width=7cm]{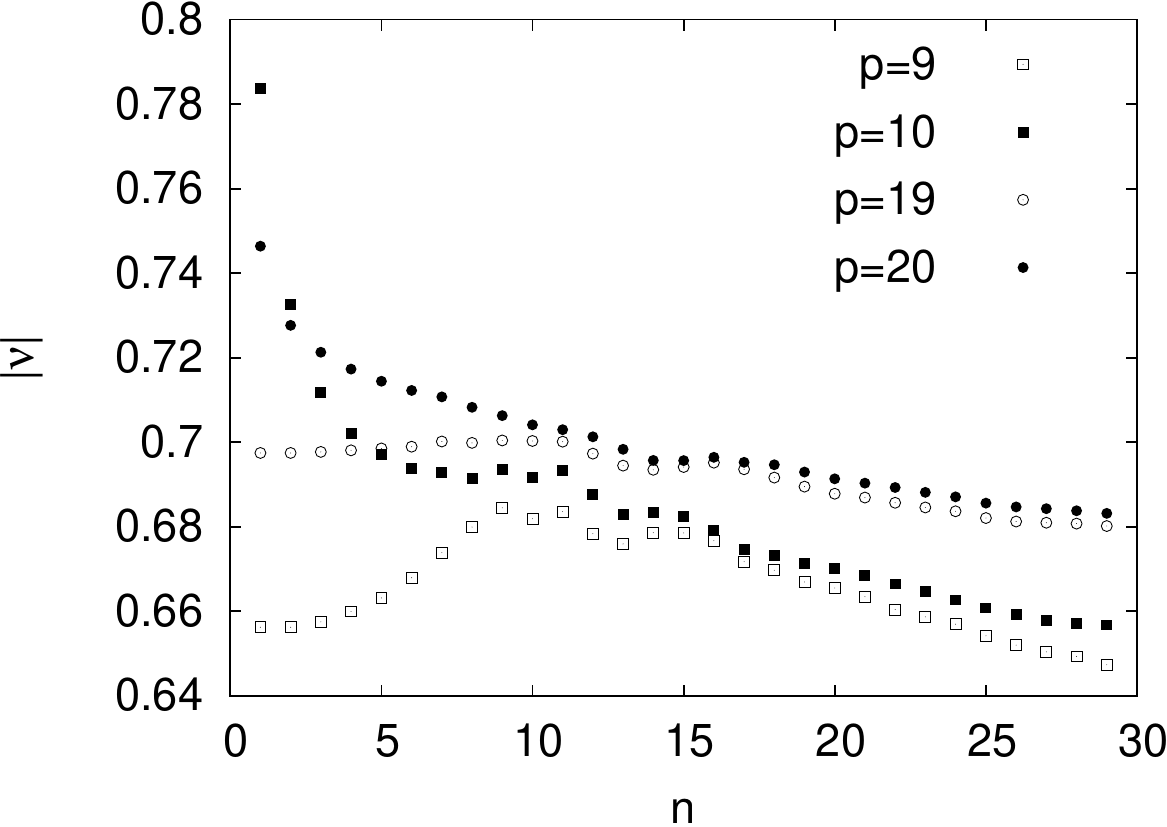}\\
{\footnotesize (c)}\hskip-5mm\includegraphics[width=7cm]{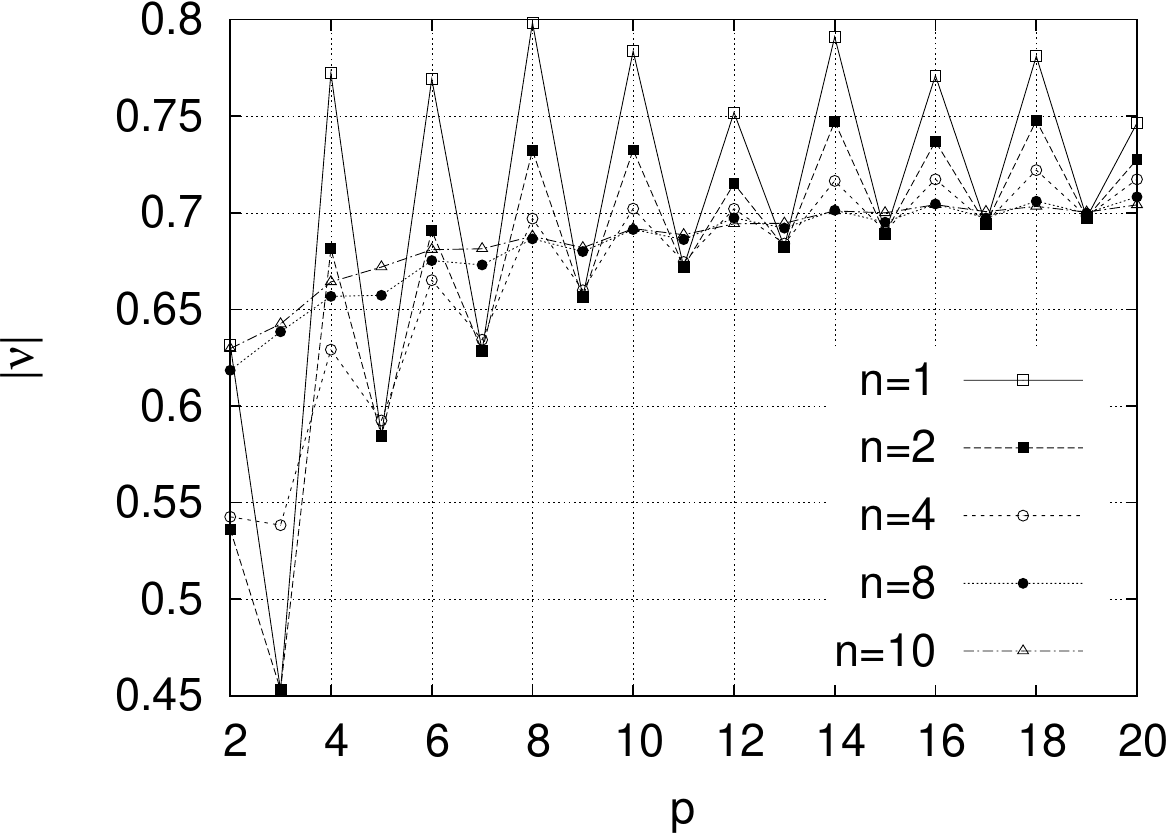}%
\hskip10mm{\footnotesize (d)}\hskip-5mm\includegraphics[width=7cm]{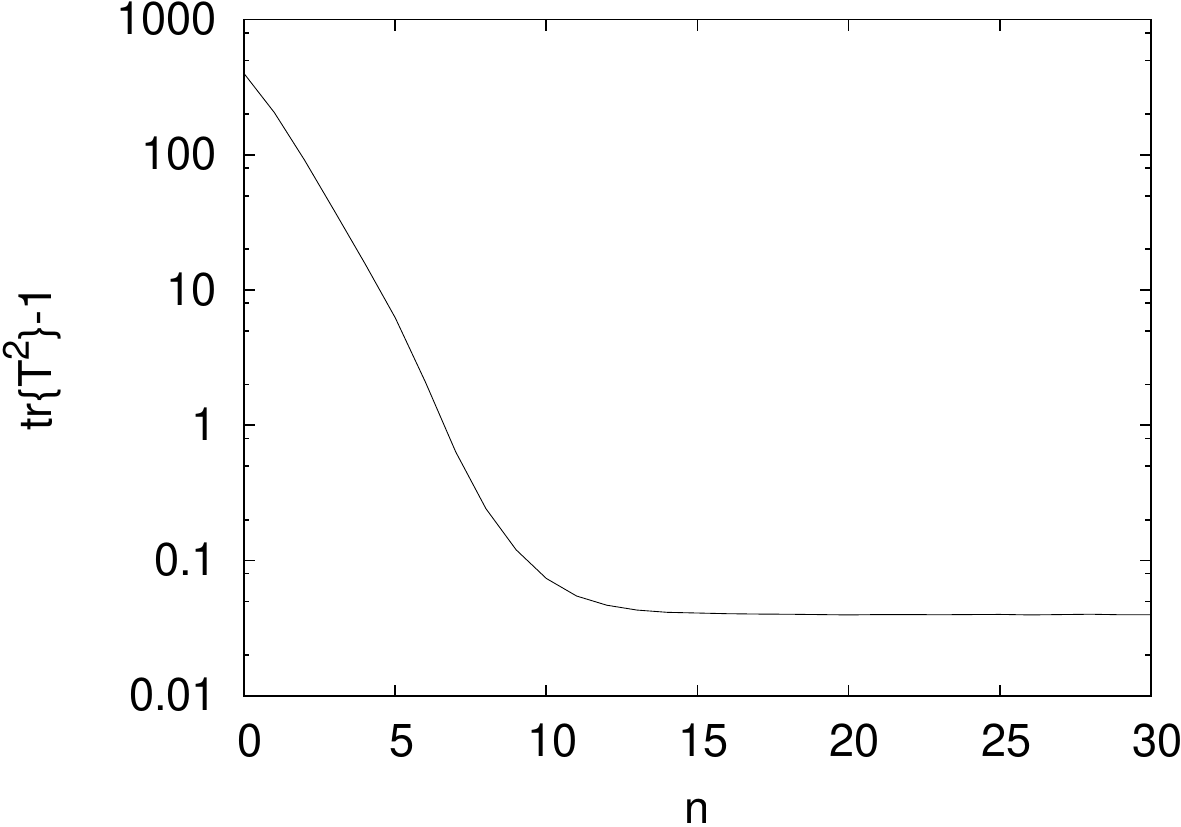}
\caption{The dependence of the leading RP resonance for the Perturbed cat map system at $K=0.9$ on the trajectory length $n$ and the number of base functions spanned along one axis $p$, using the hybrid method with $\gamma=1$ (a)-(c), and the evolution of the trace $\Tr(T^{(t)} T^{(t)\dag}$ for $p=20$ (d). The overlap integrals represented by the matrix elements $T^{(t)}_{i,j}$ are calculated by extrapolating their Simpson's rule approximants \cite{atkinson} on $1000\times1000$, $2000\times 2000$ and $4000\times 4000$ grids of points.}
\label{pic:sp_hyb_pm_K-0.9}
\end{figure}
In figure \ref{pic:sp_hyb_pm_K-0.9} we show the results of our approach for the case of $K=0.9$, where the system is already close to being marginally stable. We show the absolute value of the sub-dominant eigenvalue as a function of the basis size $p$ and the number of timesteps $n$ taken. The case of $n=1$ is equivalent to a direct diagonalization approach, whereas the case of small $p$ corresponds more closely (but is not equivalent) to the linear predictor methods. As we can see in figure \ref{pic:sp_hyb_pm_K-0.9}(a), if we only increase $p$ while keeping $n=1$ (standard diagonalization), or if we keep $p=1$ and increase $n$, the convergence is poor. Good convergence is only obtained by increasing both $p$ and $n$ simultaneously, which is also the main idea behind the hybrid approach. Figures \ref{pic:sp_hyb_pm_K-0.9}(b) and \ref{pic:sp_hyb_pm_K-0.9}(c) demonstrate the convergence properties when taking either a constant $p$ (b) or constant $n$ (c) cross section. It is interesting to note that direct diagonalisation ($n=1$) is very sensitive to the parity of the basis size $p$, but that the hybrid method tends to even out these differences with increasing $n$.\par
Figure \ref{pic:sp_hyb_pm_K-0.9}(b) also shows a significant change in convergence for roughly $n>10$. In figure \ref{pic:sp_hyb_pm_K-0.9}(d), the weight as given in equation (\ref{eq:weight}) is shown. In an exact calculation this is expected to drop to $0$, however, due to the noise present in the correlation computation, the number saturates at a higher level. The time $n$ at which the saturation is reached is also the cutoff point for the hybrid calculation, as is reflected in the convergence properties of figure \ref{pic:sp_hyb_pm_K-0.9}(b).\par
\begin{figure}
\centering
\includegraphics[width=10cm]{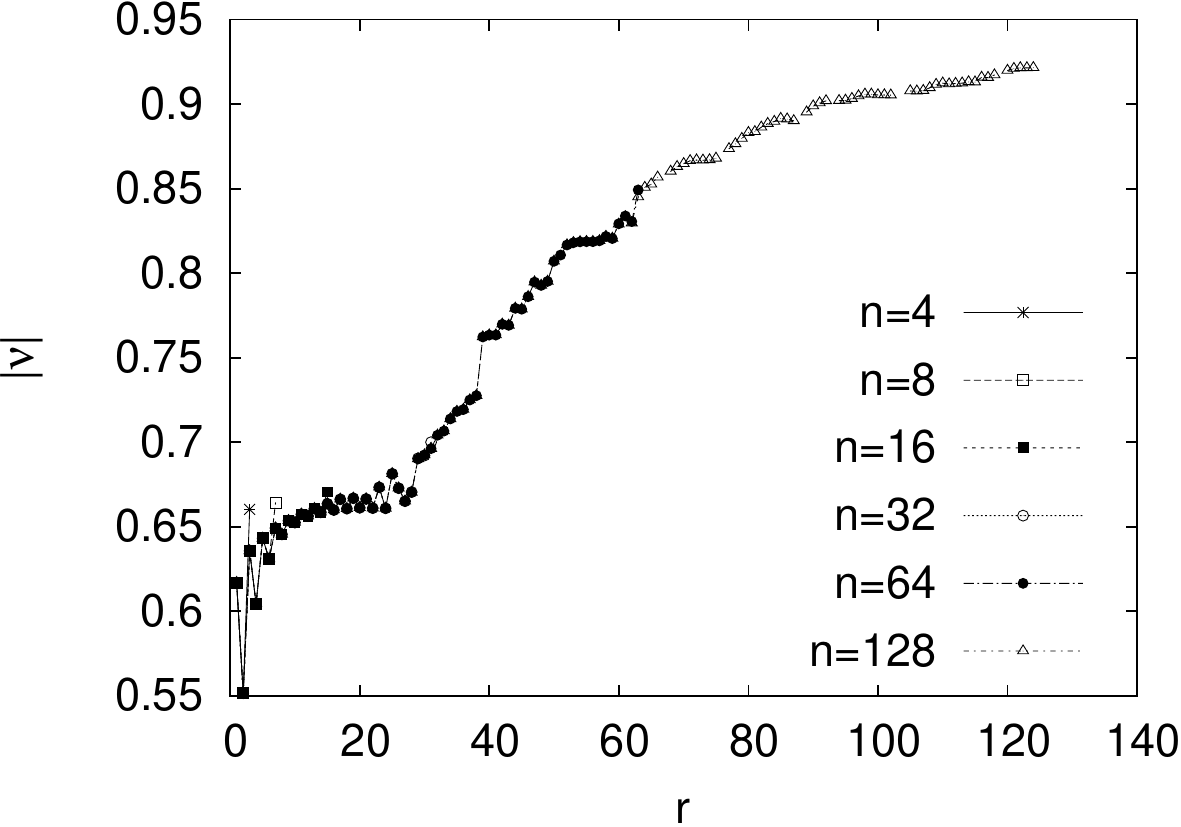}
\caption{The leading RP resonance as a function of auto-correlation signal $C(t) = \mu(g \cdot L^t \circ g)$ of length $n$ and considered resonances $r$ obtained using the linear predictor from the auto-correlation of an observable $g(x, y) = A [\sin(\pi x^2)\cos(2\pi y) - \sin(2\pi x) \cos(2\pi y^2)]$ in the perturbed cat map system for $K=0.9$ 
with $A$ chosen so that $\int_{[0,1)^2} \dd x\, \dd y\, g(x,y)^2 = 1$. The $C(t)$ is calculated with the absolute precision $3\cdot 10^{-6}$ by averaging the time auto-correlations $C(t,\xi) = \frac{1}{n-t} \sum_{i=0}^{n-t-1} g(\xi) g(f^{-t}(\xi))$ of length $n=1024$ over $10^8$ initial points uniformly distributed on the torus $\xi\in[0,1)^2$.}
\label{pic:sp_metd}
\end{figure}

In figure \ref{pic:sp_metd} we give the largest resonance as obtained via the linear predictor method for an observable chosen to not have any of the system symmetries. We can see that the procedure converges when increasing the number of chosen resonances $r$, until a certain value is reached, beyond which its value tends towards the unit circle. This is representative of the phenomenon as explained in section \ref{sec:lp} and is a serious limitation of the single variable linear prediction methods.

\begin{figure}
\centering
\includegraphics[width=9cm]{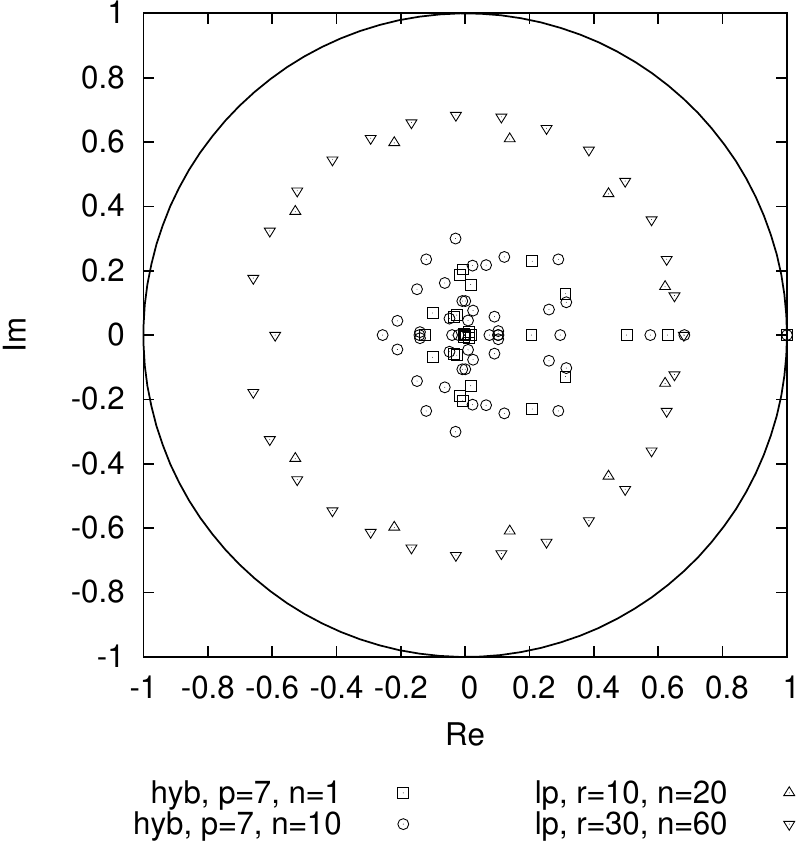}
\caption{The spectrum of resonances obtained using the hybrid method and the LP approach for different time intervals $[0,n]$ and the number of considered resonances $r$ in LP, or basis size along one axis $p$ in the hybrid method for the perturbed cat map system with $K=0.9$. For the details on numerical calculation of matrix elements $T_{i,j}^{(t)}$ and used auto-correlations see the captions under figures \ref{pic:sp_metd} and \ref{pic:sp_hyb_pm_K-0.9}.}
\label{pic:spectrums}
\end{figure}

In figure \ref{pic:spectrums} we show a comparison of full resonance spectra obtained via linear predictors in the least square approach and the hybrid method proposed here. We see that using the direct diagonalization ($n=1$) and the hybrid method with $n=10$ they give essentially different spectra for same number of spectral points given $p^2$, with the direct diagonalization tending to an underestimation of the main resonances. The spectrum obtained via LP captures the main resonance well and agrees with the hybrid method, however even for $r=10$ resonances one can observe that it is only the main resonance that is attainable by the LP approach, as all the other calculated resonances that roughly form a circle  are representative of the phenomenon described in section \ref{sec:lp}.

\begin{figure}
\centering
\includegraphics[width=9cm]{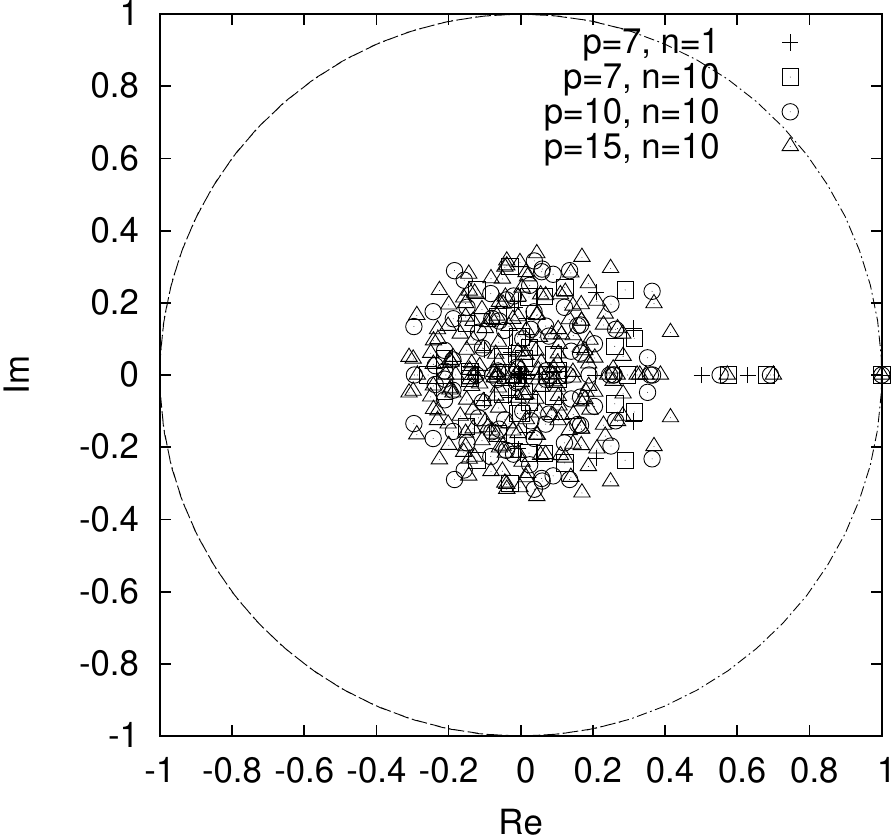}
\caption{The spectrum of resonances obtained using the hybrid method for different basis sizes along one axis $p$ for the perturbed cat map system with $K=0.9$. The direct diagonalization approach ($n=1$) for $p=7$ is shown as crosses for comparison. For the details on numerical calculation of matrix elements $T_{i,j}^{(t)}$ and auto-correlations used see the captions under figure \ref{pic:sp_hyb_pm_K-0.9}.}
\label{pic:spectrums_cmp}
\end{figure}

The convergence properties of the hybrid method for the full spectrum as a function of the basis size are demonstrated in figure \ref{pic:spectrums_cmp}. It can be seen that for all values of $p=7,10,15$, the leading eigenvalues already appear to have converged quite well for $n=10$, this being the natural cutoff time as explained earlier. For all these cases, there also exists a cloud of eigenvalues around the origin with a roughly constant radius that likely corresponds to the essential spectrum of the PF operator. Its radius slowly grows with $p$, which is believed to be due to the noise in the matrix element computations, a phenomenon not entirely unlike the one observed in the LP approach, but which does not affect the calculations in a critical manner. For comparison, the direct diagonalization results are also given.

\begin{figure}[!htb]
\centering
{\footnotesize (a)}\hskip-5mm\includegraphics[width=7cm]{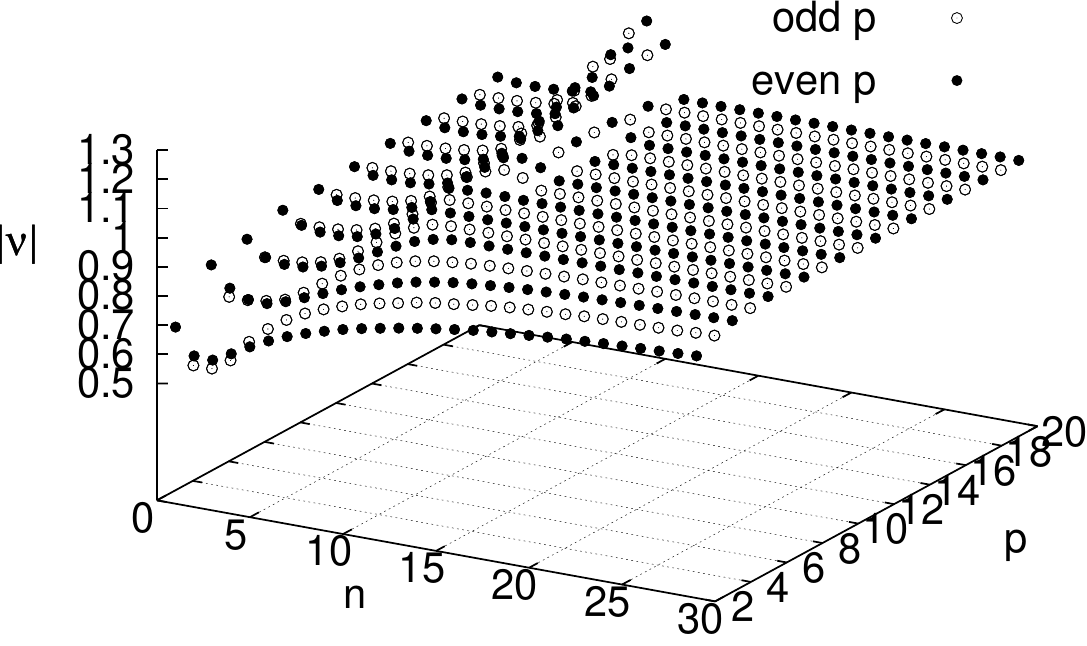}%
\hskip10mm{\footnotesize  (b)}\hskip-5mm\includegraphics[width=7cm]{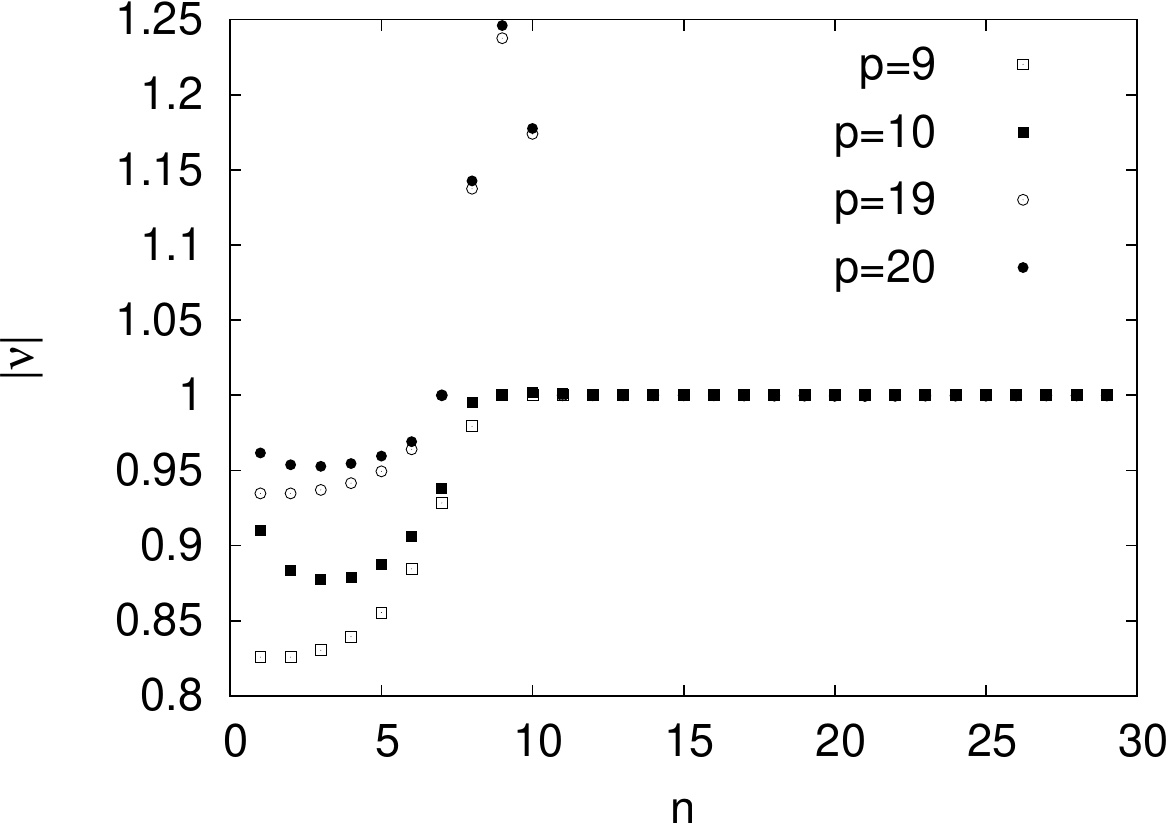}\\
{\footnotesize (c)}\hskip-5mm\includegraphics[width=7cm]{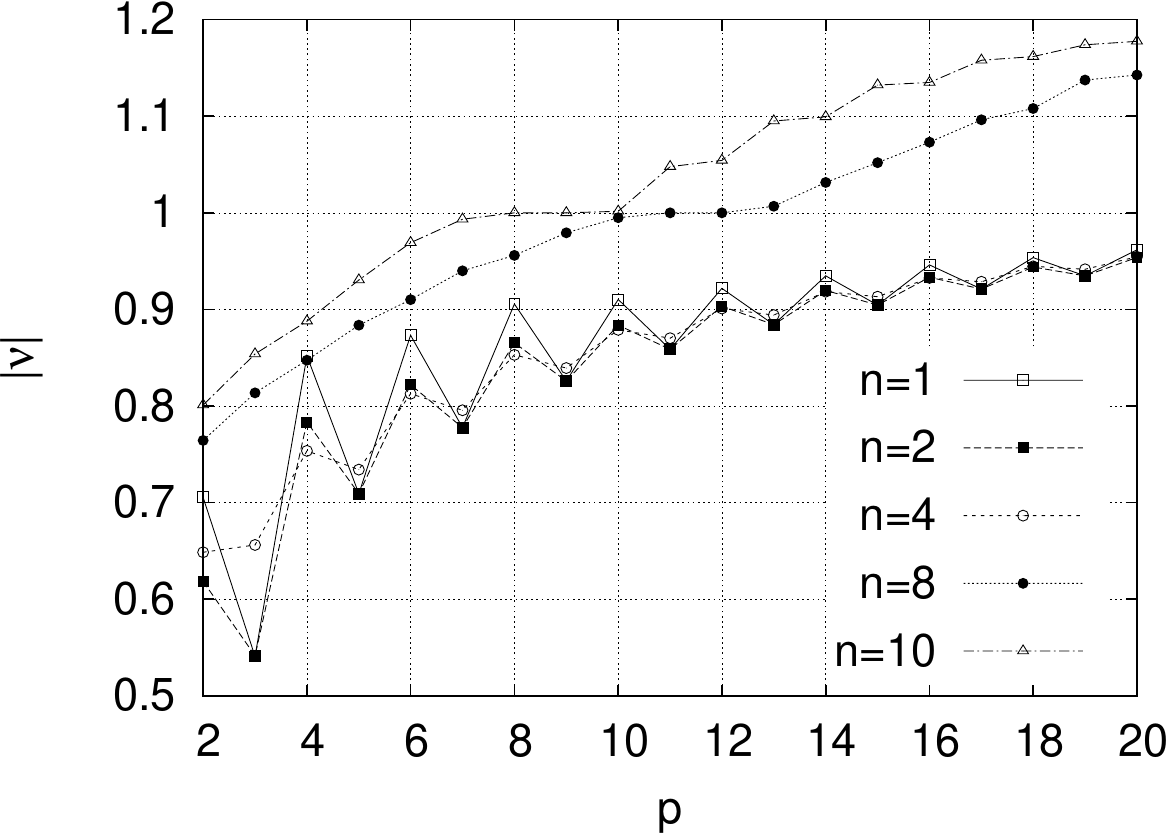}%
\hskip10mm{\footnotesize  (d)}\hskip-5mm\includegraphics[width=7cm]{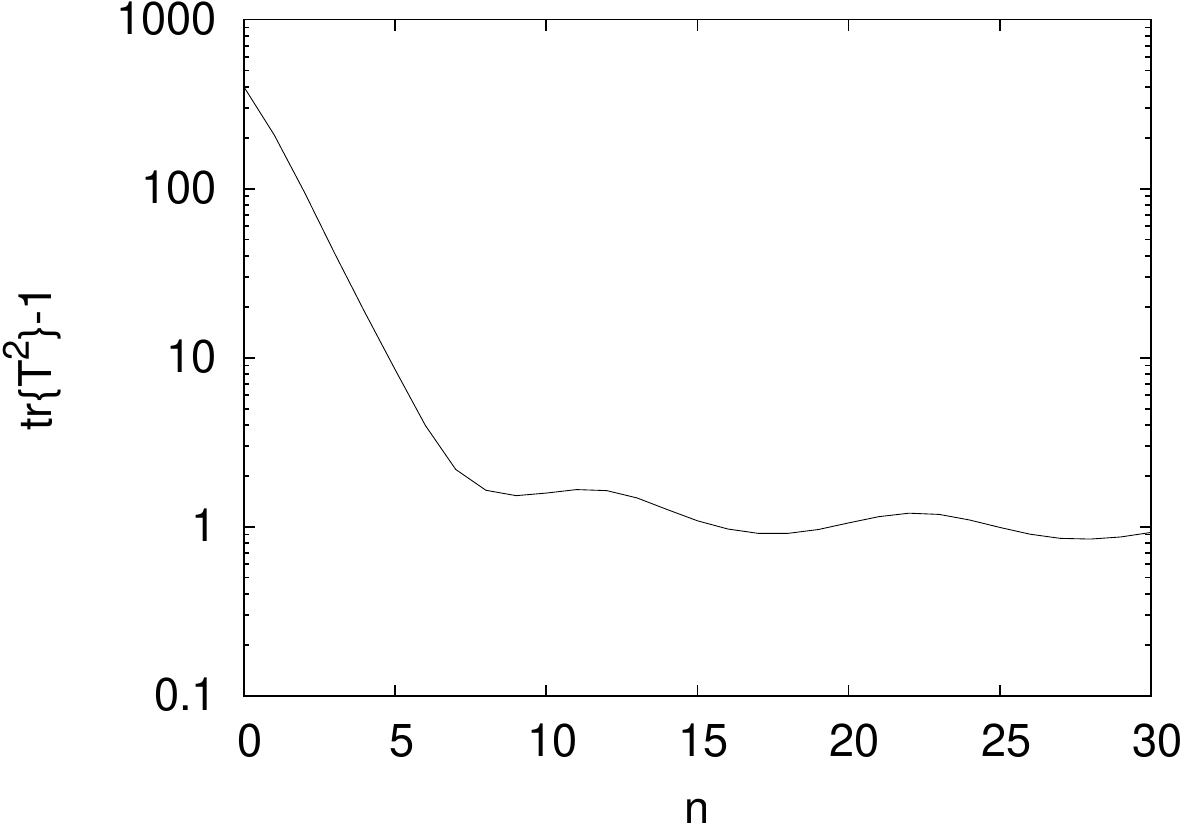}
\caption{
The leading RP resonance of the perturbed cat map at $K=1.1$ as a function of the trajectory length $n$ and the number of base functions spanned along one axis $p$ using the hybrid method with $\gamma=1$ (a)-(c) and the evolution of the trace $\Tr(T^{(t)\, 2})$ for $p=20$. Numerical details of calculations are in the caption under figure \ref{pic:sp_hyb_pm_K-0.9}.}
\label{pic:sp_hyb_pm_K-1.1}
\end{figure}
As an interesting example, we may consider the case of the perturbed cat map for $K=1.1$. At this parameter, the map ceases to be ergodic and attains regular islands. The "resonances" corresponding to the regular islands should have their absolute values at unity. Figure \ref{pic:sp_hyb_pm_K-1.1} repeats the calculation of figure \ref{pic:sp_hyb_pm_K-0.9} for such a system. We may see that the hybrid method easily captures these "resonances" close to unity, whereas both direct diagonalization as well as linear predictors (not shown) struggle to reach this value and underestimate the "resonances". One interesting failure of the method can be seen for large $p$ at around $n=10$, where the largest eigenvalue is overestimated and is pushed outside the unit circle. As can be seen in figure \ref{pic:sp_hyb_pm_K-1.1}(d), the weight $w_n$ never drops towards $0$ but maintains a plateau after $n>10$ due to the regular islands. It is only after this time that the initially larger chaotic signal decays and the small regular islands are then properly resolved in the correlation functions. Around $n\approx10$ there exists a competition between the chaotic signal contributions and the regular ones, and this is the likely reason why the hybrid method fails for when terminating the calculation at precisely these $n$. 

\section{Conclusions}

We have presented a new numerical method to calculate the RP resonances in dynamical systems with compact phase-space, which is particularly useful when one is limited to a small number of observables. The method is a hybrid between the linear prediction and diagonalization of the coarse-grained Perron-Frobenius operator. It considers many simultaneous observables, which is the main advantage of the direct diagonalization technique, over many time steps, which is the advantage of the linear prediction methods. It also requires far less observables for a good convergence than the direct diagonalization, and provides its own criterion as to the number of time steps that give best results. It is shown in the perturbed cat map system that the presented method, in the contrast to other mentioned approaches, gives a convergent and unambiguous value for the RP resonances.

\section*{Acknowledgements}

MH gratefully acknowledges the financial support by Slovenian Research Agency, grant Z1-0875-1554-08. The authors would like thank Toma\v z Prosen for interest in the work, encouraging discussions and useful comments.

\end{document}